# Kinetic Processes and surfactant design of Group I elements on CZTS ($\bar{1}\bar{1}\bar{2}$) surface

*Kejie Bao[1], Haolin Liu[1], Kinfai Tse[1], Chunlei Yang[2], Guohua Zhong[2], Junyi Zhu[1,*]*

[1]Department of Physics, The Chinese University of Hong Kong, Shatin

[2]Shenzhen Institutes of Advanced Technology, Chinese Academy of Sciences

ABSTRACT

$Cu_2ZnSnS_4$ (CZTS) is a promising thin-film solar-cell material consisted of earth abundant and nontoxic elements. Yet, there exists a fundamental bottle neck that hinders the performance of the device due to complexed intrinsic defects properties and detrimental secondary phases. Recently, it was proven experimentally that Na and K in co-evaporation growth of CZTS can enlarge the grain size and suppress formation of ZnS secondary phase near surface, but the reasons are not well understood. We used first principle calculations to investigate the kinetic processes on CZTS ($\bar{1}\bar{1}\bar{2}$) surface involving Group I elements, including Na, K, and Cs, to demonstrate their surfactant effects. Both the structure of the reconstructed surfaces involving Group I elements and various diffusion paths of a Zn ad-atom in these reconstructed surfaces were explored. The advantages and concerns of the surfactant effects of Na, K, Cs, were systematically compared and discussed. Although Group I elements protect Cu sites on the subsurface layer, a disordered metastable configuration with a diffusion barrier of about 400meV was found. Therefore, a precise control of growth condition is essential to avoid the metastable phase. In addition, our studies provide a systematical design principle for surfactant effects during the growth.

1. INTRODUCTION

Thin film of $Cu_2ZnSnS_4$ (CZTS), as a solar-cell absorber material, has attracted vast research interest in past years due to several benefits[1–10]. With all non-toxic and earth-abundant elements and ideal[11,12] band gap about 1.5eV[13], CZTS is an excellent candidate for next-generation photovoltaic technology[14–18]. Recently, the efficiency of pure CZTS solar cell and CZTS(Se) alloy has reached 11%[3] and 12.6%[4], respectively. However, the efficiency of CZTS solar cell is limited due to large concentration fluctuation of constituent elements[6,19,20]. For example, defect complex of $Zn_{Cu}$ and $Cu_{Zn}$ has very low formation energy that leads to significant reductions of band gap and open circuit voltage [19,20]. Sn related anti-sites are usually considered as deep centers that hinder the efficiencies[19,20]. Recently, it was also found that during the surface growth, Zn atoms will occupy sublayer Cu sites spontaneously and form Zn rich secondary phases that may serve as nonradiative recombination centers[21].

Surface modification method can be effective to solve these problems[22–24]. Surfactants, which are mainly used in the epitaxial growth of semiconductors, can be applied to modify surface thermodynamics and kinetics[25–33]. Different surfactant atoms including alkali atoms are proposed to change the electronic properties near the surface and reduce the possibility of Zn invading Cu site[21,34]. The surfactant design for specific materials is challenging because of complex thermodynamic and kinetic condition during the growth. **In general, the choice should follow a few guidelines: (1) low bulk incorporation and volatility to ensure the surfactant atoms stay on the surface**[25,26]**; (2) saturation of the surface dangling bonds by satisfying the electron counting model (ECM)**[35–37]**;(3) metallic nature of the surfactants relative to the bulk elements to enhance the surface diffusion**[25,26]**; (4) in CZTS case, specially, large** **surfactants should effectively protect vulnerable Cu sites in the sublayer[21]. Further, we can define a critical temperature range that satisfies all the kinetic criteria, as an important indicator of surfactant effectiveness.**

Recently, several experimental studies discovered that alkali atoms including Na, K could increase the efficiency of CZTS solar cell through enlarging the grain size and passivating non-radiative trap states[38,39,48,40–47]. Na, K, Cs might act as a surfactant and suppressor of non-radiative recombination at grain boundaries in CZTS[34,42,46]. Nevertheless, the theoretical investigations are still lacking. Only one research paper[21] has discussed the kinetic process on the formation of surface $Zn_{Cu}$ defects and investigated the surfactant effect of K. It was found that K as a surfactant can effectively prevent the Zn invasion on the sublayer Cu sties. However, the surfactant effect of Na, which could be more relevant to experimental studies, has not been studied and the systematic investigation of the surfactant effect of alkali atoms has not been performed. In addition, the kinetic pathway investigated in the previous literature is very limited. Therefore, there could easily be undiscovered surfactant related surface configurations and kinetic pathways due to the intrinsic complexity in this material system.

To answer these complex problems, it's necessary to review the current knowledge of the intrinsic surface reconstructions. XRD patterns indicate the most preferred surface orientation are $(112)/(\bar{1}\bar{1}\bar{2})$ polar surfaces during the growth of CZTS[49–52]. One previous work[12] explored the stability and electronic properties of these polar surfaces and revealed that Cu-depleted defects could stabilize the surfaces in experimental growth condition. $Zn_{Cu}$ defect and $V_{Cu}$ defect are the most stable defects formed on anion and cation terminated surfaces[12], respectively. In this paper, we will mainly focus on investigations of the general surfactant effects of group I elements and search for the most effective surfactants. In our investigation, anion-terminated ($\bar{1}\bar{1}2$) surface will be studied because they are the most stable ones under the most preferred experimental conditions[50]. To understand the physics of the Zn diffusion with surfactants, based on the guidelines of surfactants design, we (1) explored the thermodynamic and kinetic processes of the bulk incorporation of Group I (Na, K, Cs) elements; (2) searched for the most stable surface reconstructions with alkali atoms involved; (3) calculated the diffusion paths and energy barriers on top surface; (4) discovered a new metastable configuration that was missed in previous literature and calculated the diffusion barriers for Zn invading process; (5) evaluated and discussed the advantages and concerns of Group I elements (Na, K, Cs).

## 2. METHODOLOGIES

The calculations of bulks and slabs were based on density functional theory[53,54] as implemented in Vienna Ab-initio Simulation Package (VASP)[55,56], within the Perdew-Burke-Ernzerhopf (PBE) generalized-gradient approximation (GGA)[57]. Plane waves[58,59] with a kinetic energy cutoff of 400 eV were used as basis sets. Bulk defect properties were obtained using a supercell approach, where single defect was introduced in a 64-atom cell with (3×3×3) Monkhorst-Pack[60] $k$-point sampling, consistent with previous method[47]. For surface calculations, (2×1) unit cell slabs containing 6 bi-layers of CZTS and at least 20 Å vacuum were used. (3×3×1) Monkhorst-Pack $k$-point mesh was used for integration over Brillouin zone. Pseudo-hydrogen atoms with charge q = 1.75e, q = 1.5e, and q = e were added on the bottom surface to passivate dangling bonds of Cu, Zn, and Sn atoms, respectively, following the scheme in previous literature[34]. The atoms of the four bottom bi-layers were kept fixed, while all other atoms were free to move during optimization. Electronic relaxation of surface calculations is converged to 10⁻⁵ eV. Ionic relaxation of surface calculations is converged until all atomic force are less than 0.01 eV/ Å. Diffusion path and transition state energy were determined by the climbing image nudged elastic band (CI-NEB) method[61]. The formation energy of defects is calculated as following[62]:

$$E^f[X] = E_{\text{tot}}[X] - E_{\text{tot}}[\text{bulk}] - \sum_i n_i \mu_i \quad (1)$$

where $E_{\text{tot}}[X]$ is the total energy derived from a supercell calculation containing the defect X, and $E_{tot}[\text{bulk}]$ is the total energy for host crystal using the equivalent supercell. The integer $n_i$ represents the number of host atoms or impurity atoms that have been added to or removed from the supercell, and the $\mu_i$ is the corresponding chemical potentials of these atoms. The formation energy of the reconstructed surface is calculated as following[12,34]:

$$\sigma_{\text{sur}} = \frac{1}{A_{sur}} \left[ E_{sur} - \sum_i n_i \mu_i - \sum_j n_j \mu_j^{PCP} \right] \quad (2)$$

where $A_{sur}$ is the area of the surface, $E_{sur}$ is the total energy of the CZTS 6-bilayer slabs. $\mu_j^{PCP}$ is pseudo chemical potential of the bottom pseudo-hydrogen[34,63].

## 3. RESULTS AND DISCUSSIONS

In this section, the properties of various surfactants of Group I (Na, K, Cs) elements were systematically investigated and compared through the aforementioned aspects: volatility, bulk incorporation, stability, surface diffusion enhancement, vulnerable sites protection and growth temperature window. Although volatility is important in surfactant design, the boiling points of Na and K are similar and above 1000 K[64]. However, that of Cs is slightly lower[64] and could be more volatile.

3.1 Bulk incorporation

Bulk incorporation of Group I (Na, K, Cs) were calculated thermodynamically and kinetically since it is a prerequisite of surfactants. In the experiments, CZTS is often grown by sputtering[50] or co-evaporation[52] methods, which are often considered as highly non-equilibrium growth methods. So, it is necessary to include kinetic analysis. Moreover, the thermodynamic analysis can still be a valid approximation to evaluate the surfactant effects because near the growth front, there exists thermodynamic driving force towards the local equilibrium configuration as long as the growth speed is not too rapid. Therefore, for thermodynamics, the formation energy of surfactants related bulk defects were calculated, as shown in Figure 1; and for kinetics, the energy barriers for single surfactant atom diffusing from surface to the bulk were calculated, as shown in Figure 2 and Figure 3. In Figure 1, chemical potential points A~H were adopted from our previous article[34], where A~D correspond to Cu poor limit and E~H correspond to Cu rich limit.

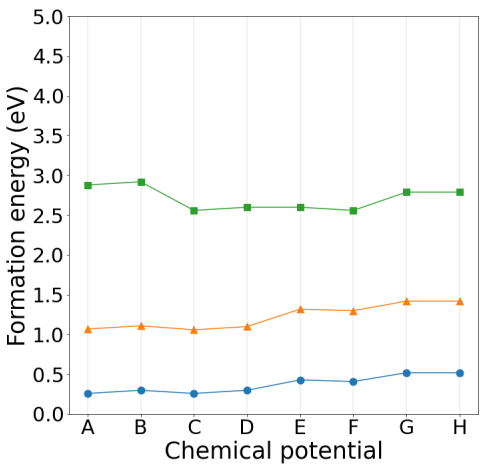
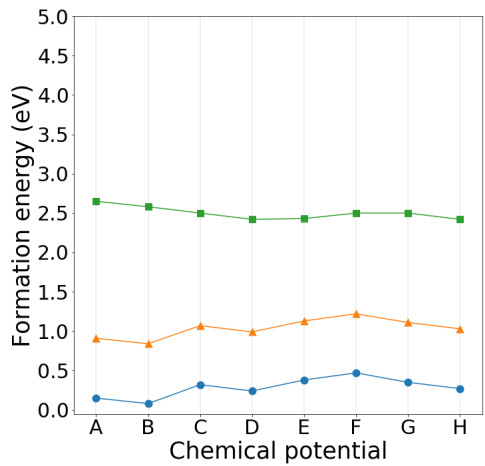

(a). $Na_{Cu}$, $K_{Cu}$ and $Cs_{Cu}$            (b). $Na_{Zn}$, $K_{Zn}$ and $Cs_{Zn}$

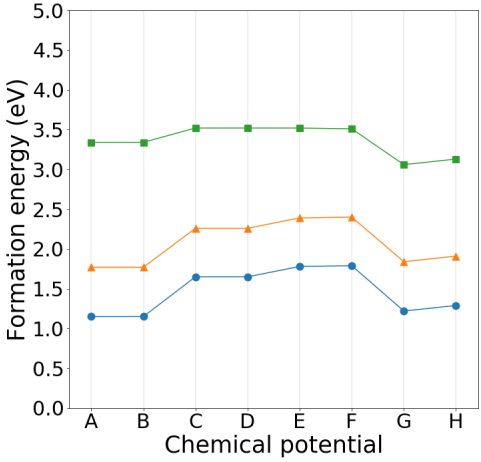 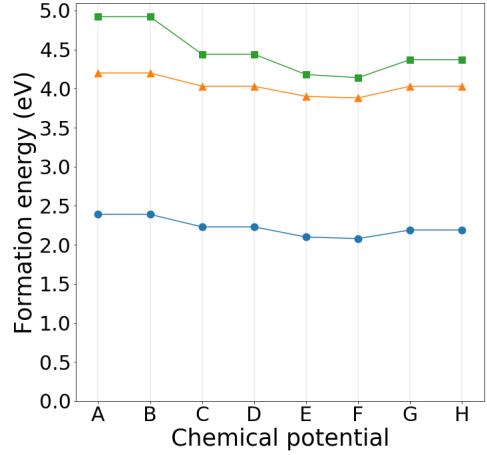

(c). Na$_{Sn}$, K$_{Sn}$ and Cs$_{Sn}$  (d). Na$_i$, K$_i$ and Cs$_i$

**Figure 1.** Formation energy (in units of eV) of Group I (Na, K, Cs) elements related substitutional (a: Cu, b: Zn, c: Sn) and interstitial (d) point defects in CZTS bulk. Na, K, Cs are represented by blue, orange, and green line, respectively. Chemical potential points A~H are adopted from our previous article[34], where A~D correspond to Cu poor limit and E~H correspond to Cu rich limit.

After reproducing the calculations of Na and K in previous work[47], the point defects for Cs were further calculated. Chemical potential of surfactant elements (Na, K, Cs) were assumed in the rich limit. As shown in Figure 1, three types of substitutional point defects: dopant$_{Cu}$, dopant$_{Zn}$, dopant$_{Sn}$, and interstitial point defects, dopant$_i$ (all possible sites were checked and only sites with lowest energy were plotted here) were studied, where dopants stand for Na, K, and Cs. The formation energy of dopant$_{Cu}$ and dopant$_{Zn}$ are close and low because alkali atoms could reduce the strain energy by occupying Cu or Zn sites indicating they are most common point defects in CZTS bulk consistent with former results[47]. The formation energy of Cs related substitutional and interstitial are highest as expected because the size of Cs is larger than that of Na and K. Moreover, if we assume a Boltzmann distribution at typical CZTS growth temperature 500°C[8], concentration ratio between Na and K, and that between Na and Cs defects are $10^{-8}$ and $10^{-24}$, respectively, indicating much lower K, Cs concentration than Na.

For kinetics, CI-NEB calculations were performed to find the energy barriers for Group I elements diffusing from the surface to the subsurface. First, the most stable initial configurations of single element attachment on CZTS ($\bar{1}\bar{1}\bar{2}$) surface were searched. We found that Na, K, and Cs ad-atoms all favor the hollow site, as shown in Figure 2. The final configurations were searched as the surfactant atoms completely incorporate into the cation layer at subsurface. Our diffusion calculations indicate that Cs induces unstable surface structures. Therefore, Cs atom is unlikely to incorporate into the CZTS bulk due to the huge size of Cs. The minimum energy paths (MEPs) and the initial, saddle, final configurations for Na and K were shown in Figure 3.

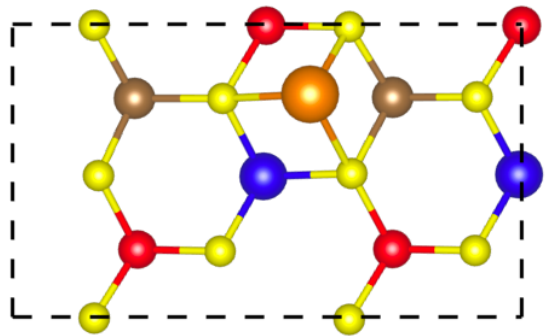

**Figure 2.** The most stable configuration of CZTS ($\bar{1}\bar{1}\bar{2}$) surface with single Group I (Na, K, Cs) atom. Only the top three layers are shown here. The red, brown, blue, yellow, and orange balls represent Cu, Zn, Sn, S and general Group I (Na, K, Cs) atoms, respectively.

The diffusion barriers for Na and K are about 1.5eV and 2.2eV, respectively, indicating the diffusion processes for both Na and K are slow. Although the formation energy of Na related defects is relatively low in our defect calculation, the diffusion process is still unfavorable due to the high diffusion barrier. Therefore, low concentration of Na in CZTS bulk should be expected. Also, Na related stable point defects are generally benign[10], thus Na is still a reasonable choice for the surfactants in CZTS even if a small amount of Na is incorporated into the bulk. In summary, Group I (Na, K, Cs) all satisfy the prerequisite of low bulk incorporation, under the consideration of both thermodynamics and kinetics.

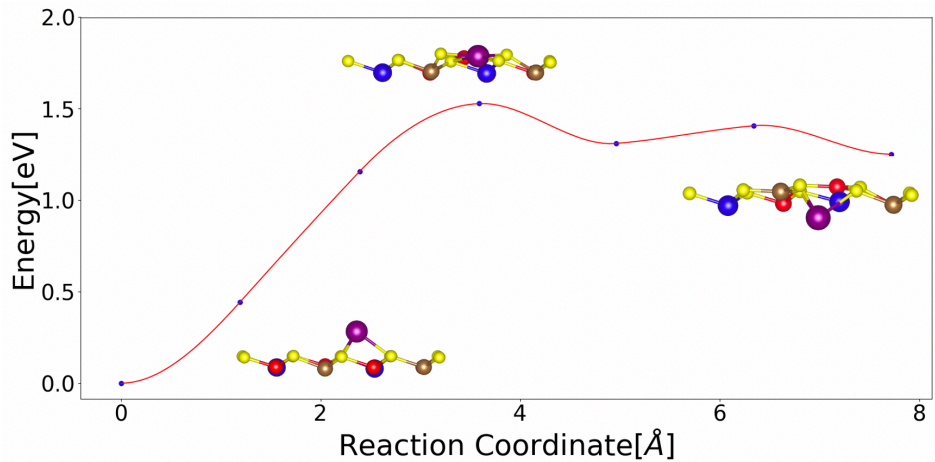

(a) Na diffusion

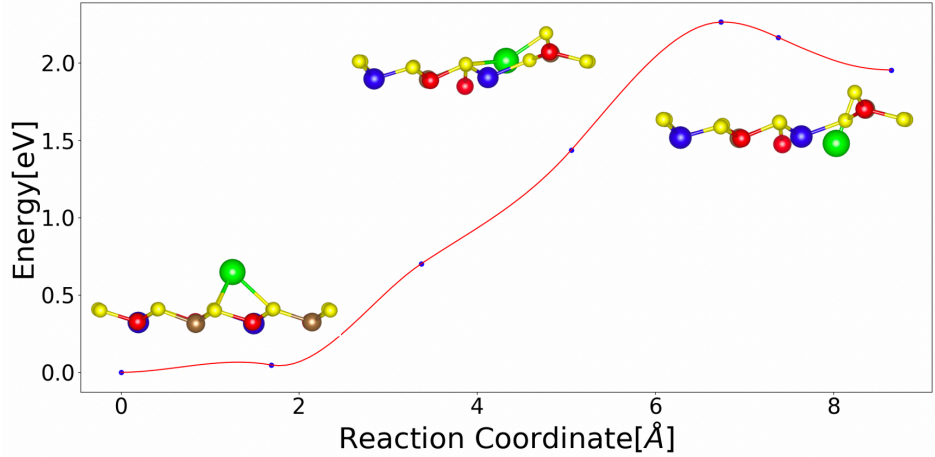

(b) K diffusion

**Figure 3.** Diffusion process for single Na and K atom from the surface incorporating into the sublayer. Initial, saddle and final configurations are indicated. Na and K are represented by purple and green balls, respectively. Cs induces unstable surface structures from our calculations.

3.2 Stability

Group I (Na, K, Cs) atoms, as surfactants, significantly change the electronic environment near ($\bar{1}\bar{1}\bar{2}$) surface through saturating the dangling bonds, according to our calculation details. Various alkali atoms related surface reconstructions were considered here, and the most stable configurations were used for surface diffusion calculations. Based on the ECM, totally four Group I atoms were put on the (2×1) CZTS ($\bar{1}\bar{1}\bar{2}$) surface to fully passivate all dangling bonds because four electrons are needed for bare surface. We calculated totally thirty types of configurations for each group I element and found that five representative configurations, including H3 (hollow) rectangle (Figure 4(a)), T4 (top) rectangle (Figure 4(b)), zigzag (Figure 4(c)), mixed rectangle (Figure 4(d)), and mixed parallelogram (Figure 4(e)) were relatively stable and presented here. All three Group I elements share the same most stable configuration, a mixed parallelogram, with two H3 sites and two T4 sites, as shown in Figure 4(e). Surfactant atoms tend to be separated to reduce the surface stress.

Additionally, we calculated the formation energy of the most stable configuration for each group I element to illustrate the relative stability at different chemical potential points, as shown in Figure 5. Similar with previous defects calculations, Group I elements (Na, K, Cs) are all assumed at rich limit. The formation energy for Na and K do not change much at different chemical potential points, however, that of Cs changes dramatically from Cu poor limit to Cu rich limit. Experimentally, the typical growth condition is at Cu poor limit (A~D). We found that the formation energy of K-capped reconstructed surface is lowest at this regime among the surfactant elements. Noted, even if surfactant atoms are not in the rich limit, our conclusion still holds if the concentration of the surfactant candidates are similar in the experiments.

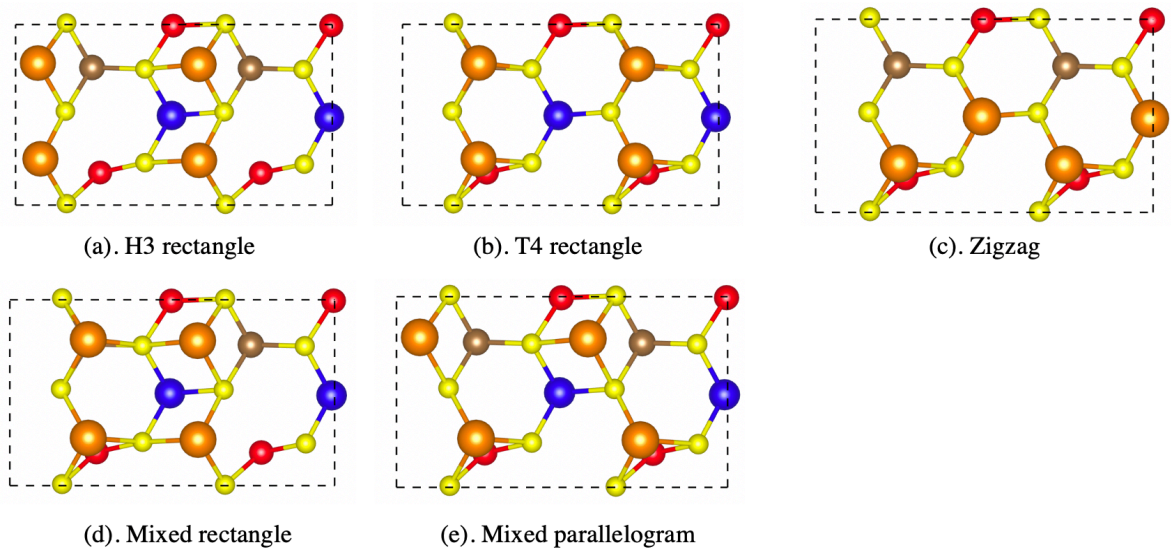

(a). H3 rectangle  (b). T4 rectangle  (c). Zigzag
(d). Mixed rectangle  (e). Mixed parallelogram

**Figure 4.** Some possible configurations that satisfy electron counting model (ECM) including (a) H3 rectangle, (b) T4 rectangle, (c) Zigzag, (d) Mixed rectangle, and (e) Mixed parallelogram configurations. Atomic symbols follow Figure 2.

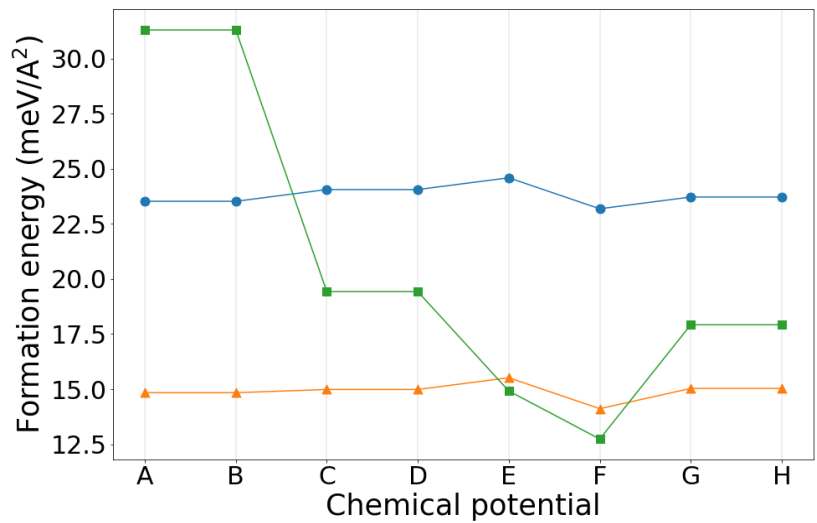

**Figure 5.** Formation energy of the Group I (Na, K, Cs) atoms related reconstructed surface in different chemical potential points. Na, K, Cs are represented by circle (blue), triangle (orange) and square (green), respectively.

3.3 Surface diffusion enhancement

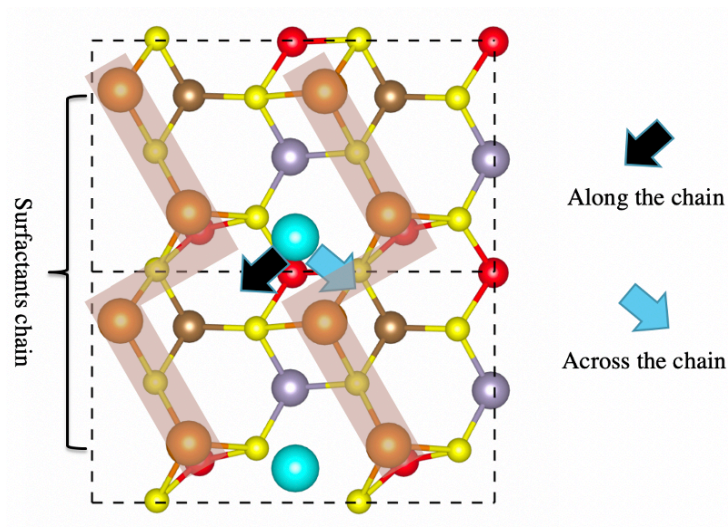

**Figure 6.** Shadowed area schematically illustrates the zigzag chain of Group I (Na, K, Cs) surfactants. Two possible diffusion paths of single Zn ad-atom along the chain and cross chain are illustrated by the black and grey arrows, respectively. Atomic symbols follow Figure 2. Zn ad-atom is represented by cyan ball.

---

Group I elements (Na, K, Cs) are expected to enhance the surface diffusion due to their metallic property. Since invasion behavior of Zn against Cu is problematic in CZTS, we investigate the surface diffusion of Zn as an illustration. The diffusion paths and energy barriers of single Zn ad-atom on the reconstructed surface were calculated in this section. As shown in figure 6, the alkali atoms form a zigzag chain in the mixed parallelogram configuration. Although the diffusion paths on the surface are numerous, we adopted paths starting from most stable surface adsorption site of Zn with surfactants involved. There are two orthogonal paths: along the zigzag chain and across the zigzag chain.

We have calculated the MEPs for single Zn ad-atom diffusing along and across the surfactant chain of Na, K, Cs, as shown in Figure 7 and Figure 8, respectively. Our calculated energy barriers of K were comparable with previous literature[21]. For diffusion along the zigzag chain, Na, K, and Cs share the similar diffusion paths. The saddle point of Zn ad-atom tends to stay in the center of the parallelogram, as shown in Figure 7. The energy barriers for diffusion along the zigzag chain for Na, K, Cs are 0.104 eV, 0.310 eV, 0.357 eV, respectively. The differences of the energy barriers here are mainly determined by the size effect of alkali atoms. Larger size corresponds to higher energy cost of the diffusion here.

For diffusion across the zigzag chain, Na has a completely different diffusion path from the K and Cs since the initial position of Zn ad-atom is higher than Na-capped surface while lower than K-capped or Cs-capped surface. The barriers for diffusion across the zigzag chain for Na, K, Cs are 0.250 eV, 0.260 eV, 0.310 eV, respectively. Moreover, there is only one saddle point for the Na case while there are two saddle points for the K and Cs case according our calculation, as shown in Figure 8. For the Na case, the Zn ad-atom is higher than the Na capped surface, thus Zn ad-atom tends to cross the zigzag chain through the absorption site located on top of the surface without distorting the original surface structure that may cost extra energy. However, different from the Na capped surface, Zn ad-atom on K and Cs capped surfaces prefer to diffuse under the surfactant plane despite causing larger distortion than the Na case. The distortion can be demonstrated by the surfactant atom of K or Cs with a symbol of "A" marked in Figure 8(b) and Figure 8(c) on the diffusion pathway of Zn. A concerted movement of both Zn and A can be observed. The "A" atom moves laterally from the original top site to a hollow site when Zn ad-atom attempts to cross the surfactant chain. Then, "A" atom will gradually recover to its original position. Asymmetry of the right and left sides of the zigzag chain results in two saddle points for this diffusion path. Size effect is also important to determine the energy barriers.

Through the former analysis, the diffusion barriers on Na capped surface are the lowest, indicating Na significantly enhances Zn diffusion and prevents Zn atoms accumulating at the local minimal on the surface. Since Na is less metallic than K and Cs, disruption to the diffusion from the Na ions should be considerably weaker than K and Cs cases, leading to lower diffusion barriers. In summary, Na has the best effect to enhance the diffusion on top surface among three surfactant candidates.

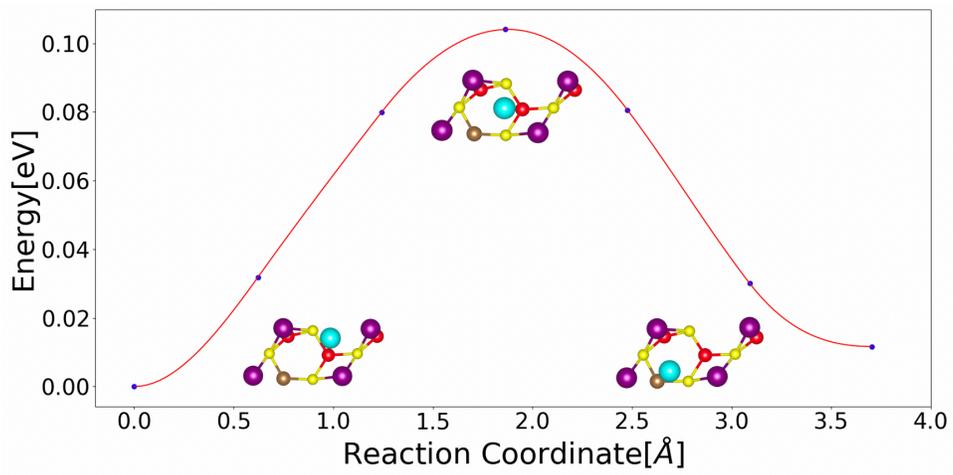

(a) Na capped surface

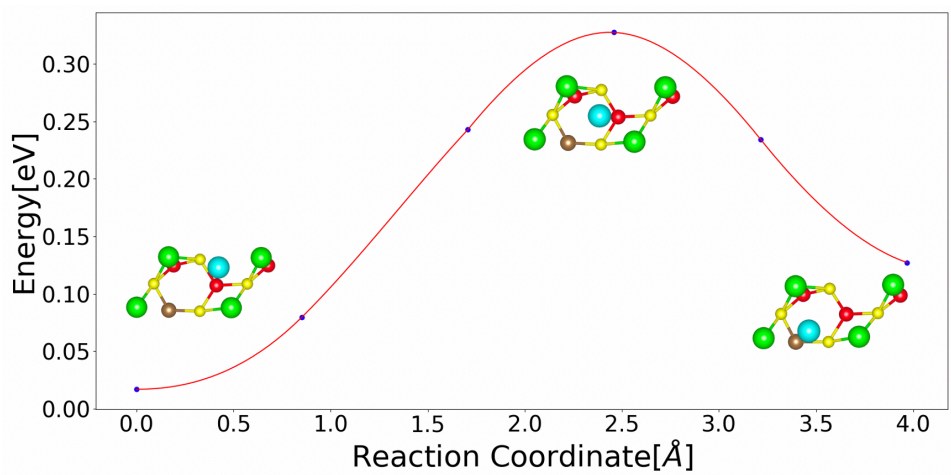

(b) K capped surface

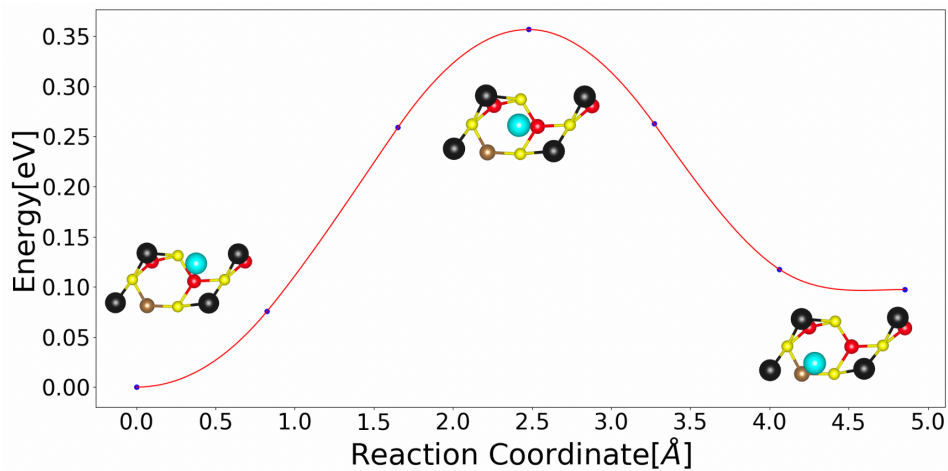

(c) Cs capped surface

**Figure 7.** The MEPs of single Zn ad-atom diffusion along the (a) Na, (b) K, (c) Cs zigzag chain. Initial, saddle, and final configurations are indicated. Na, K, Cs, and Zn ad-atom are represented by purple, green, black, and cyan balls, respectively.

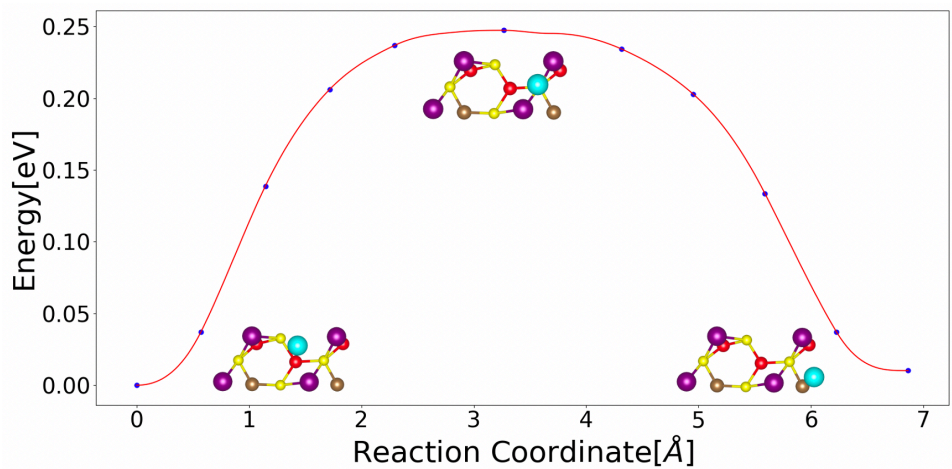

(a) Na capped surface

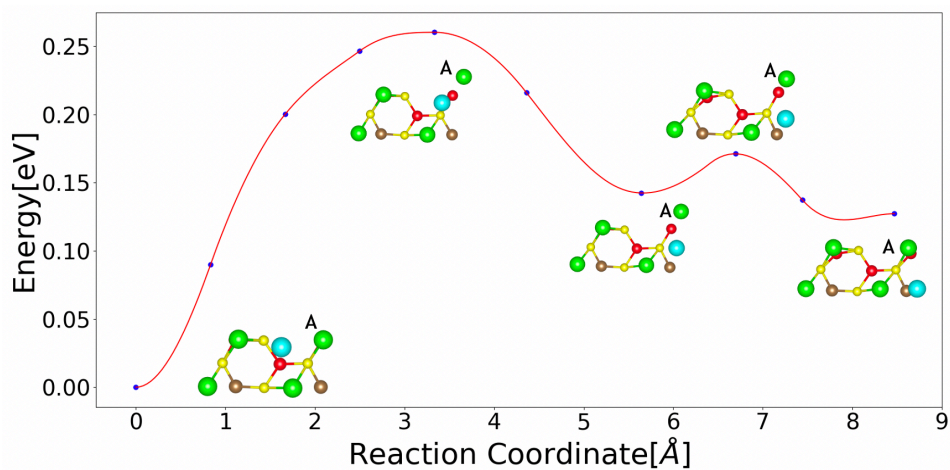

(b) K capped surface

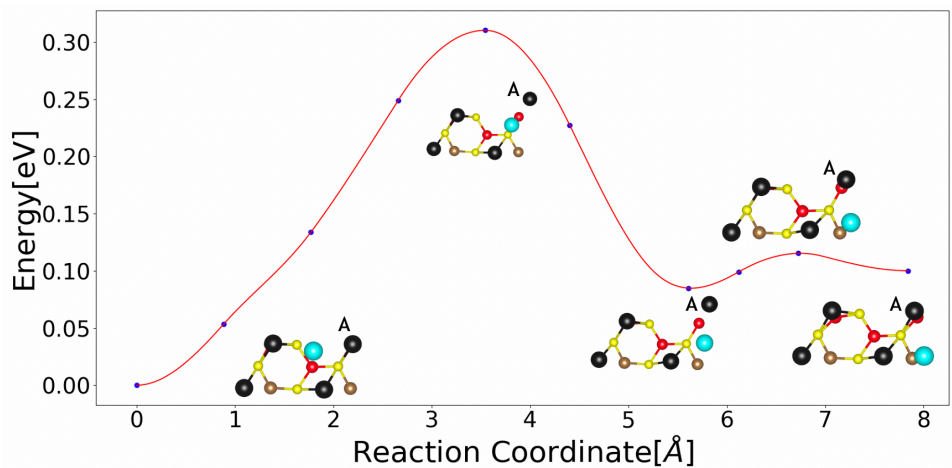

(c). Cs capped surface

**Figure 8.** The MEPs of single Zn ad-atom diffusion across the (a) Na, (b) K, (c) Cs zigzag chain. Initial, saddle and final configurations are indicated. Atomic symbols follow Figure 7. Distorted K or Cs atoms due to Zn ad-atom diffusion is labeled by "A" in (b) and (c).

3.4 Vulnerable sites protection

To investigate which candidate has the best protection for the Cu atoms on the subsurface layers, we performed further calculations in this section. Due to the systematic complexity and large interstitial space near the surface layers, there could be various local minima. During our diffusion investigations, we indeed discovered new metastable configurations that may trap Zn on the subsurface layer on Na, K, and Cs capped surfaces. For the Na case, one S atom is pushed up by Zn ad-atom incorporation with the breaking of two Cu-S bonds and one Sn-S bond, as shown in Figure 9. As a result, the Sn atom changed from a four-fold coordination to a three-fold one. Next, two extra electrons from Zn ad-atom transfer to the Sn atom and form a lone pair. Therefore, this metastable configuration satisfies ECM and the total energy is largely reduced.

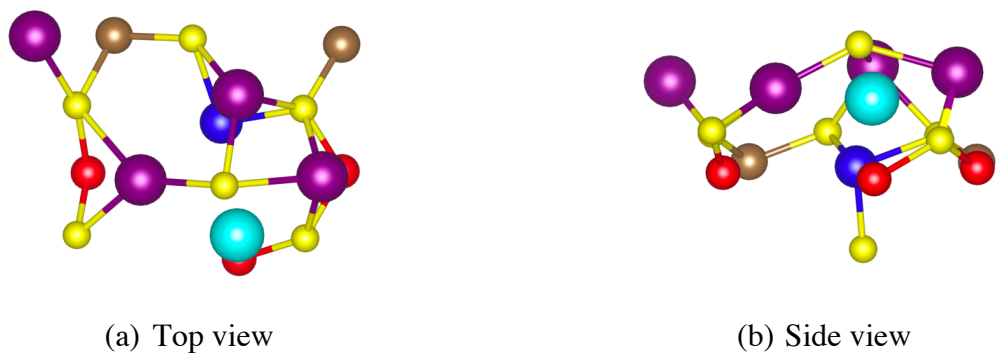

(a) Top view  (b) Side view

**Figure 9.** Trapped configuration during the diffusion process for Zn ad-atom incorporating from the surface into the sublayer for Na case. The red, brown, blue, yellow, purple, and cyan balls represent Cu, Zn, Sn, S, Na and Zn ad-atom, respectively.

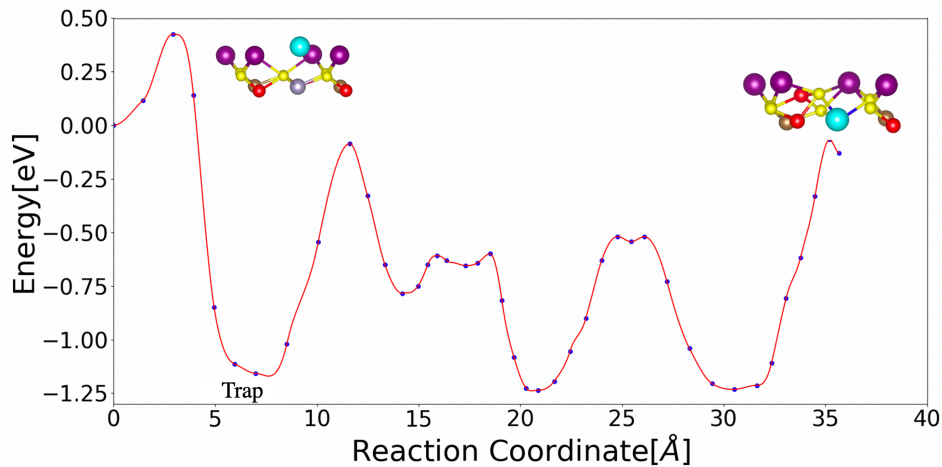

**Figure 10.** The complete MEP of single Zn ad-atom diffusion incorporating into the underlying Cu site on the cation layer in Na-capped surface. Initial and final configurations are indicated. Atomic symbols follow Figure 9. The energy related to the trapped configuration is labelled by "Trap".

Based on this trapped configuration, a complete MEP of single Zn ad-atom incorporating into underlying Cu site at cation layer from Na-capped surface was calculated, as shown in Figure 10. This complicated diffusion path contains multiple saddle points because of the large lattice distortion with the incorporation of Zn ad-atom. After overcoming the first energy barrier, Zn ad-atom falls into a trap and the energy needed to escape from the trap is over 1 eV. Despite Zn ad-atom has low probability to invade vulnerable Cu site on Na passivated surface, it faces a severe trapping problem with a large distortion. We further calculated the similar trapping configuration for K and Cs capped surface and found the similar local minima, as shown in Figure 11. The energy barriers that prevent the falling into the trapped configuration for Na, K, Cs capped surface are 0.410 eV, 0.348 eV and 0.489 eV, respectively. Surprisingly, we observed a non-monotonic trend with the energy barrier for K as the lowest among these three elements.

Therefore, we can use the first energy barrier that prevents the trapping process as an indicator to quantify the protection ability of the Group I elements.

There are several factors including size effect and distortion argument to explain this abnormal phenomenon. Cs provides better protection due to its largest size and strongest ionic potential among the three elements. Moreover, for K case, smaller distortion contributes to the reduction of the total energy at the saddle point. Actually, the bond length of K and S (2.92Å), and that of Zn and S (2.44Å) near the saddle point are both very close to that in the $K_2S$ and ZnS compound[65]. Therefore, the local stress at the saddle point is largely relaxed. If Boltzmann distribution at a typical growth temperature of 500°C [8] is assumed, the percentage of trapped configuration is about 0.21%, 0.52% and 0.06% for Na, K, Cs case, respectively. In summary, all three elements are reasonably good and Cs has the best protection ability of vulnerable Cu sites from Zn invasion among three surfactant candidates.

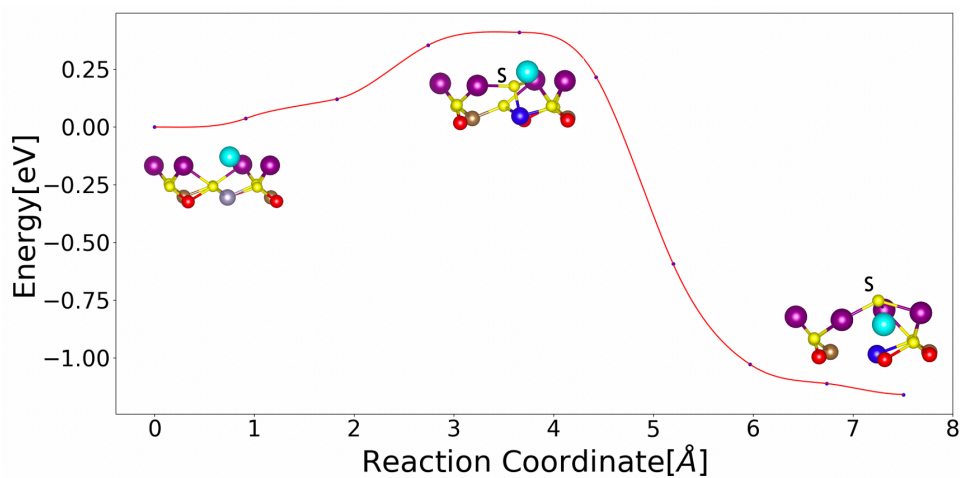

(a) Na capped surface

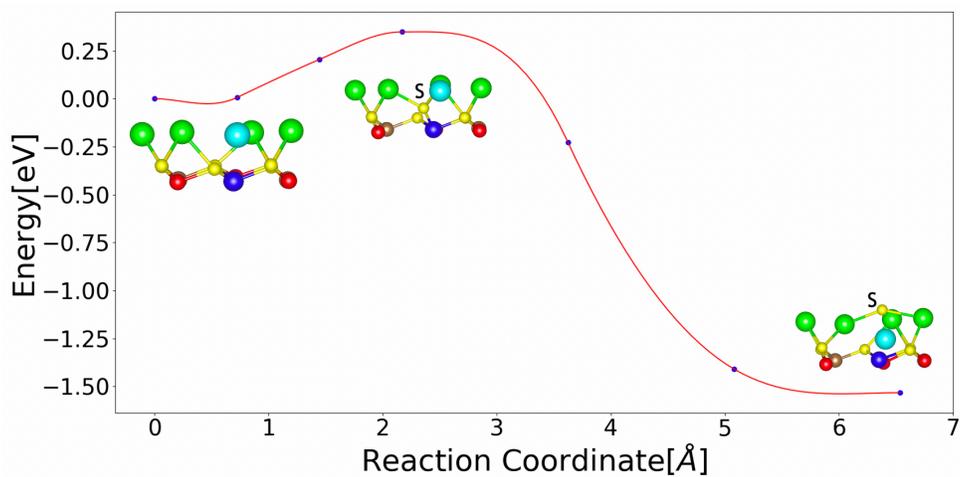

(b) K capped surface

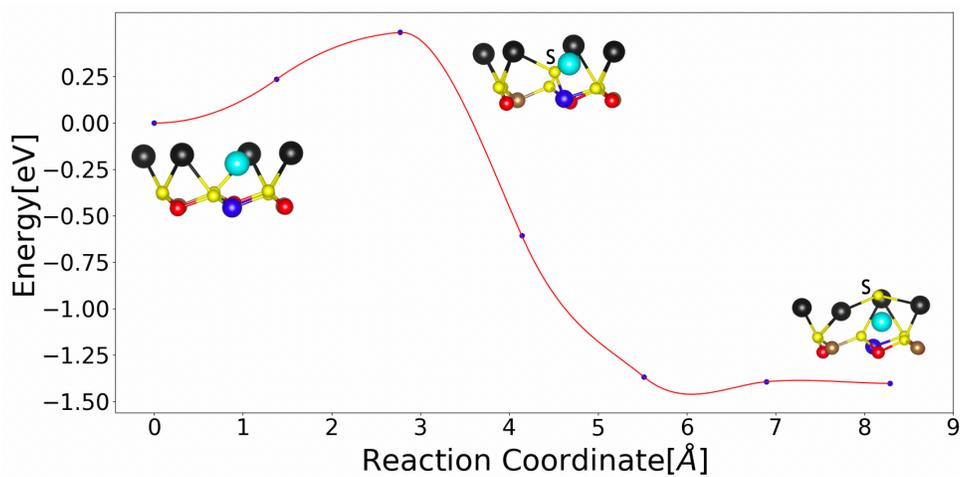

(c) Cs capped surface

**Figure 11.** The MEPs of single Zn ad-atom diffusion from the surface to the trapped intermediate configuration for (a) Na, (b) K, (c) Cs case. Initial, saddle and final configurations are indicated. Atomic symbols follow Figure 7. The moving up S atom is labelled by "S".

3.5 Comparison

The advantages and concerns of Group I elements (Na, K, Cs), as surfactants, were systematically compared. This comparative study is based on the above analysis in six aspects: volatility, bulk incorporation, stability, surface diffusion enhancement, vulnerable sites protection, and growth temperature window, as defined by the difference between the energy barriers on surface diffusion and that of the diffusion into the sublayer, as shown in Table 1. We employed a three-color spotlight approach to depict good, fair, and limited properties or functionalities of the surfactants in blue, green and red colors, respectively, as shown in Table 2. The assessments on the 6 aspects are based on the following facts: (1) The boiling points for Na and K are both above 1000K while Cs is slightly lower; (2) The formation energy of dopant related defects and the dopant incorporating kinetic barriers for Cs are the highest, followed by K and Na; (3) Surfactant involved reconstructed surface for K is the most stable, followed by Na and Cs; (4) The surface diffusion barriers of Zn ad-atom are the lowest for Na and the highest for Cs; (5) The incorporation barrier of Zn ad-atom is the highest for Cs and the lowest for K; (6) Growth temperature window is the largest for Na and smallest for K. To evaluate the overall performance, we assign 3 points for good aspect, 2 points for fair aspect, and 1 point for limited aspect, and sum the scores for each surfactant. Based on this simple evaluation system, Na is the best surfactant among three candidates.

**Table 1**. **Energy barriers with a unit of eV of different Zn diffusion path types: along the zigzag chain, across the zigzag chain, and incorporation into the sublayer for Na, K, Cs capped surface.** The last row shows the growth temperature window, as defined by the difference between the energy barriers on surface diffusion and that of the diffusion into the sublayer.

| Path type | Na-capped | K-capped | Cs-capped |
|---|---|---|---|
| **On top surface** (Along the zigzag chain) | 0.104 | 0.310 | 0.357 |
| **On top surface** (Across the zigzag chain) | 0.250 | 0.260 | 0.310 |
| **Incorporation into the sublayer** | 0.410 | 0.348 | 0.489 |
| **Growth temperature window** | 0.160 | 0.038 | 0.132 |

**Table 2. Qualitative evaluation of surfactant effects.**

| Surfactant related properties | Na | K | Cs |
|---|---|---|---|
| 1. Volatility | Good | Good | Fair |
| 2. Bulk incorporation | Fair | Fair | Good |
| 3. Stability | Fair | Good | Fair |
| 4. Surface diffusion enhancement | Good | Fair | Limited |
| 5. Vulnerable sites protection | Fair | Limited | Good |
| 6. Growth temperature window | Good | Limited | Fair |
| Total | Good | Fair | Fair |

legends

| Level | Colour |
|---|---|
| Good | 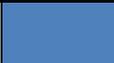 |
| Fair | 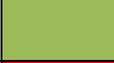 |
| Limited | 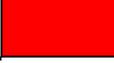 |

## 4 CONCLUSION

In conclusion, we systematically investigated the surfactant related properties and performance of Group I elements including Na, K, Cs and evaluated them in six aspects: volatility, bulk incorporation, stability, surface diffusion enhancement, vulnerable sites protection, and growth temperature window. Based on our evaluation system, we concluded that Na is the best surfactant among these three alkali elements mainly because it can largely enhance the surface diffusion and have a relatively wide growth temperature window. Moreover, we discovered the low energy trapped configuration, which can be detrimental during the CZTS growth. Our works demonstrate a standard procedure on the surfactant design in terms of surface kinetics during the growth. Our evaluation system could be revised and extended to other material systems on the predictions of surfactant effects.

AUTHOR INFORMATION

**Corresponding Author**

*e-mail: jyzhu@phy.cuhk.edu.hk

ACKNOWLEDGMENT

Computing resources were provided by the High-Performance Cluster Computing Centre, the Chinese University of Hong Kong (CUHK). This work was supported by the start-up funding at CUHK and CAS-CUHK joint lab. Financial support of General Research Fund (Grant No. 14301318, 14319416) from Research Grants Council in Hong Kong is gratefully acknowledged.

REFERENCES

(1) Todorov, T. K.; Tang, J.; Bag, S.; Gunawan, O.; Gokmen, T.; Zhu, Y.; Mitzi, D. B. Beyond 11% Efficiency: Characteristics of State-of-the-Art Cu2ZnSn (S, Se) 4 Solar Cells. *Adv. Energy Mater.* **2013**, *3* (1), 34–38.

(2) Mitzi, D. B.; Gunawan, O.; Todorov, T. K.; Wang, K.; Guha, S. The Path towards a High-Performance Solution-Processed Kesterite Solar Cell. *Sol. Energy Mater. Sol. Cells* **2011**, *95* (6), 1421–1436.

(3) Yan, C.; Huang, J.; Sun, K.; Johnston, S.; Zhang, Y.; Sun, H.; Pu, A.; He, M.; Liu, F.; Eder, K.; others. Cu 2 ZnSnS 4 Solar Cells with over 10% Power Conversion Efficiency Enabled by Heterojunction Heat Treatment. *Nat. Energy* **2018**, *3* (9), 764.

(4) Wang, W.; Winkler, M. T.; Gunawan, O.; Gokmen, T.; Todorov, T. K.; Zhu, Y.; Mitzi, D. B. Device Characteristics of CZTSSe Thin-Film Solar Cells with 12.6% Efficiency. *Adv. Energy Mater.* **2014**, *4* (7), 1301465.

(5) Walsh, A.; Chen, S.; Wei, S.-H.; Gong, X.-G. Kesterite Thin-Film Solar Cells: Advances in Materials Modelling of Cu2ZnSnS4. *Adv. Energy Mater.* **2012**, *2* (4), 400–409.

(6) Wang, H. Progress in Thin Film Solar Cells Based On. *Int. J. Photoenergy* **2011**, *2011*.

(7) Zhu, J.; Liu, F.; Scarpulla, M. A. Strain Tuning of Native Defect Populations: The Case of Cu2ZnSn (S, Se) 4. *APL Mater.* **2014**, *2* (1), 12110.

(8) Ravindiran, M.; Praveenkumar, C. Status Review and the Future Prospects of CZTS Based Solar Cell--A Novel Approach on the Device Structure and Material Modeling for CZTS Based Photovoltaic Device. *Renew. Sustain. Energy Rev.* **2018**, *94*, 317–329.

(9)  Steinhagen, C.; Panthani, M. G.; Akhavan, V.; Goodfellow, B.; Koo, B.; Korgel, B. A. Synthesis of Cu2ZnSnS4 Nanocrystals for Use in Low-Cost Photovoltaics. *J. Am. Chem. Soc.* **2009**, *131* (35), 12554–12555.

(10) Redinger, A.; Hönes, K.; Fontané, X.; Izquierdo-Roca, V.; Saucedo, E.; Valle, N.; Pérez-Rodr\'\iguez, A.; Siebentritt, S. Detection of a ZnSe Secondary Phase in Coevaporated Cu 2 ZnSnSe 4 Thin Films. *Appl. Phys. Lett.* **2011**, *98* (10), 101907.

(11) Chen, S.; Walsh, A.; Gong, X.-G.; Wei, S.-H. Classification of Lattice Defects in the Kesterite Cu2ZnSnS4 and Cu2ZnSnSe4 Earth-Abundant Solar Cell Absorbers. *Adv. Mater.* **2013**, *25* (11), 1522–1539.

(12) Xu, P.; Chen, S.; Huang, B.; Xiang, H.-J.; Gong, X.-G.; Wei, S.-H. Stability and Electronic Structure of Cu 2 ZnSnS 4 Surfaces: First-Principles Study. *Phys. Rev. B* **2013**, *88* (4), 45427.

(13) Park, D.; Nam, D.; Jung, S.; An, S.; Gwak, J.; Yoon, K.; Yun, J. H.; Cheong, H. Optical Characterization of Cu2ZnSnSe4 Grown by Thermal Co-Evaporation. *Thin Solid Films* **2011**, *519* (21), 7386–7389.

(14) Song, X.; Ji, X.; Li, M.; Lin, W.; Luo, X.; Zhang, H. A Review on Development Prospect of CZTS Based Thin Film Solar Cells. *Int. J. Photoenergy* **2014**, *2014*.

(15) Meillaud, F.; Shah, A.; Droz, C.; Vallat-Sauvain, E.; Miazza, C. Efficiency Limits for Single-Junction and Tandem Solar Cells. *Sol. energy Mater. Sol. cells* **2006**, *90* (18–19), 2952–2959.

(16) Katagiri, H.; Jimbo, K.; Maw, W. S.; Oishi, K.; Yamazaki, M.; Araki, H.; Takeuchi, A. Development of CZTS-Based Thin Film Solar Cells. *Thin Solid Films* **2009**, *517* (7), 2455–2460.

(17) Green, M. A.; Hishikawa, Y.; Warta, W.; Dunlop, E. D.; Levi, D. H.; Hohl-Ebinger, J.; Ho-Baillie, A. W. H. Solar Cell Efficiency Tables (Version 50). *Prog. Photovoltaics Res. Appl.* **2017**, *25* (7), 668–676.

(18) Scragg, J. J.; Watjen, J. T.; Edoff, M.; Ericson, T.; Kubart, T.; Platzer-Björkman, C. A Detrimental Reaction at the Molybdenum Back Contact in Cu2ZnSn (S, Se) 4 Thin-Film Solar Cells. *J. Am. Chem. Soc.* **2012**, *134* (47), 19330–19333.

(19) Chen, S.; Gong, X. G.; Walsh, A.; Wei, S.-H. Defect Physics of the Kesterite Thin-Film Solar Cell Absorber Cu 2 ZnSnS 4. *Appl. Phys. Lett.* **2010**, *96* (2), 21902.

(20) Chen, S.; Yang, J.-H.; Gong, X.-G.; Walsh, A.; Wei, S.-H. Intrinsic Point Defects and Complexes in the Quaternary Kesterite Semiconductor Cu 2 ZnSnS 4. *Phys. Rev. B* **2010**, *81* (24), 245204.

(21) Zhang, Y.; Tse, K.; Xiao, X.; Zhu, J. Controlling Defects and Secondary Phases of CZTS by Surfactant Potassium. *Phys. Rev. Mater.* **2017**, *1* (4), 45403.

(22) Zhang, S. B.; Wei, S.-H. Reconstruction and Energetics of the Polar (112) and ($\bar{1}\bar{1}\bar{2}$) versus the Nonpolar (220) Surfaces of CuInSe 2. *Phys. Rev. B* **2002**, *65* (8), 81402.

(23) Hofmann, A.; Pettenkofer, C. The CuInSe2--CuIn3Se5 Defect Compound Interface: Electronic Structure and Band Alignment. *Appl. Phys. Lett.* **2012**, *101* (6), 62108.

(24) Jaffe, J. E.; Zunger, A. Defect-Induced Nonpolar-to-Polar Transition at the Surface of Chalcopyrite Semiconductors. *Phys. Rev. B* **2001**, *64* (24), 241304.

(25) Kandel, D.; Kaxiras, E. The Surfactant Effect in Semiconductor Thin Film Growth. *arXiv Prepr. cond-mat/9901177* **1999**.

(26) Kandel, D.; Kaxiras, E. Surfactant Mediated Crystal Growth of Semiconductors. *Phys. Rev. Lett.* **1995**, *75* (14), 2742.

(27) Zhu, J. Y.; Liu, F.; Stringfellow, G. B. Dual-Surfactant Effect to Enhance p-Type Doping in III-V Semiconductor Thin Films. *Phys. Rev. Lett.* **2008**, *101* (19), 196103.

(28) Zhu, J.; Liu, F.; Stringfellow, G. B. Enhanced Cation-Substituted p-Type Doping in GaP from Dual Surfactant Effects. *J. Cryst. Growth* **2010**, *312* (2), 174–179.

(29) Fetzer, C. M.; Lee, R. T.; Shurtleff, J. K.; Stringfellow, G. B.; Lee, S. M.; Seong, T. Y. The Use of a Surfactant (Sb) to Induce Triple Period Ordering in GaInP. *Appl. Phys. Lett.* **2000**, *76* (11), 1440–1442.

(30) Stringfellow, G. B.; Lee, R. T.; Fetzer, C. M.; Shurtleff, J. K.; Hsu, Y.; Jun, S. W.; Lee, S.; Seong, T. Y. Surfactant Effects of Dopants on Ordering in GaInP. *J. Electron. Mater.* **2000**, *29* (1), 134–139.

(31) Wixom, R. R.; Rieth, L. W.; Stringfellow, G. B. Sb and Bi Surfactant Effects on Homo-Epitaxy of GaAs on (001) Patterned Substrates. *J. Cryst. Growth* **2004**, *265* (3–4), 367–374.

(32) Howard, A. D.; Chapman, D. C.; Stringfellow, G. B. Effects of Surfactants Sb and Bi on the Incorporation of Zinc and Carbon in III/V Materials Grown by Organometallic Vapor-Phase Epitaxy. *J. Appl. Phys.* **2006**, *100* (4), 44904.

(33) Wixom, R. R.; Modine, N. A.; Stringfellow, G. B. Theory of Surfactant (Sb) Induced Reconstructions on InP (001). *Phys. Rev. B* **2003**, *67* (11), 115309.

(34) Wong, M.; Tse, K.; Zhu, J. New Types of CZTS {$\Sigma$3} {${$112$}$} Grain Boundaries: Algorithms to Passivation. *J. Phys. Chem. C* **2018**, *122* (14), 7759–7770.

(35) Pashley, M. D. Electron Counting Model and Its Application to Island Structures on Molecular-Beam Epitaxy Grown GaAs (001) and ZnSe (001). *Phys. Rev. B* **1989**, *40* (15), 10481.

(36) Duke, C. B. Semiconductor Surface Reconstruction: The Structural Chemistry of Two-Dimensional Surface Compounds. *Chem. Rev.* **1996**, *96* (4), 1237–1260.

(37) Chadi, D. J. Atomic Structure of GaAs (100)-(2$\times$ 1) and (2$\times$ 4) Reconstructed Surfaces. *J. Vac. Sci. Technol. A Vacuum, Surfaces, Film.* **1987**, *5* (4), 834–837.

(38) Gershon, T.; Shin, B.; Bojarczuk, N.; Hopstaken, M.; Mitzi, D. B.; Guha, S. The Role of Sodium as a Surfactant and Suppressor of Non-Radiative Recombination at Internal Surfaces in Cu2ZnSnS4. *Adv. Energy Mater.* **2015**, *5* (2), 1400849.

(39) Rey, G.; Babbe, F.; Weiss, T. P.; Elanzeery, H.; Melchiorre, M.; Valle, N.; El Adib, B.; Siebentritt, S. Post-Deposition Treatment of Cu2ZnSnSe4 with Alkalis. *Thin Solid Films* **2017**, *633*, 162–165.

(40)  Altamura, G.; Wang, M.; Choy, K.-L. Influence of Alkali Metals (Na, Li, Rb) on the Performance of Electrostatic Spray-Assisted Vapor Deposited Cu 2 ZnSn (S, Se) 4 Solar Cells. *Sci. Rep.* **2016**, *6*, 22109.

(41)  Johnson, M.; Baryshev, S. V; Thimsen, E.; Manno, M.; Zhang, X.; Veryovkin, I. V; Leighton, C.; Aydil, E. S. Alkali-Metal-Enhanced Grain Growth in Cu 2 ZnSnS 4 Thin Films. *Energy Environ. Sci.* **2014**, *7* (6), 1931–1938.

(42)  Mule, A.; Vermang, B.; Sylvester, M.; Brammertz, G.; Ranjbar, S.; Schnabel, T.; Gampa, N.; Meuris, M.; Poortmans, J. Effect of Different Alkali (Li, Na, K, Rb, Cs) Metals on Cu2ZnSnSe4 Solar Cells. *Thin Solid Films* **2017**, *633*, 156–161.

(43)  Hsieh, Y.-T.; Han, Q.; Jiang, C.; Song, T.-B.; Chen, H.; Meng, L.; Zhou, H.; Yang, Y. Efficiency Enhancement of Cu2ZnSn (S, Se) 4 Solar Cells via Alkali Metals Doping. *Adv. Energy Mater.* **2016**, *6* (7), 1502386.

(44)  Tong, Z.; Yan, C.; Su, Z.; Zeng, F.; Yang, J.; Li, Y.; Jiang, L.; Lai, Y.; Liu, F. Effects of Potassium Doping on Solution Processed Kesterite Cu2ZnSnS4 Thin Film Solar Cells. *Appl. Phys. Lett.* **2014**, *105* (22), 223903.

(45)  Zhou, H.; Song, T.-B.; Hsu, W.-C.; Luo, S.; Ye, S.; Duan, H.-S.; Hsu, C.-J.; Yang, W.; Yang, Y. Rational Defect Passivation of Cu2ZnSn (S, Se) 4 Photovoltaics with Solution-Processed Cu2ZnSnS4: Na Nanocrystals. *J. Am. Chem. Soc.* **2013**, *135* (43), 15998–16001.

(46)  Maeda, T.; Kawabata, A.; Wada, T. First-Principles Study on Alkali-Metal Effect of Li, Na, and K in Cu2ZnSnS4 and Cu2ZnSnSe4. *Phys. status solidi* **2015**, *12* (6), 631–637.

(47) Tse, K.; Wong, M.; Zhang, Y.; Zhang, J.; Scarpulla, M.; Zhu, J. Defect Properties of Na and K in Cu2ZnSnS4 from Hybrid Functional Calculation. *J. Appl. Phys.* **2018**, *124* (16), 165701.

(48) Sutter-Fella, C. M.; Stckelberger, J. A.; Hagendorfer, H.; La Mattina, F.; Kranz, L.; Nishiwaki, S.; Uhl, A. R.; Romanyuk, Y. E.; Tiwari, A. N. Sodium Assisted Sintering of Chalcogenides and Its Application to Solution Processed Cu2ZnSn (S, Se) 4 Thin Film Solar Cells. *Chem. Mater.* **2014**, *26* (3), 1420–1425.

(49) Pawar, S. M.; Pawar, B. S.; Moholkar, A. V; Choi, D. S.; Yun, J. H.; Moon, J. H.; Kolekar, S. S.; Kim, J. H. Single Step Electrosynthesis of Cu2ZnSnS4 (CZTS) Thin Films for Solar Cell Application. *Electrochim. Acta* **2010**, *55* (12), 4057–4061.

(50) Fernandes, P. A.; Salomé, P. M. P.; Da Cunha, A. F. Growth and Raman Scattering Characterization of Cu2ZnSnS4 Thin Films. *Thin Solid Films* **2009**, *517* (7), 2519–2523.

(51) Fernandes, P. A.; Salomé, P. M. P.; Da Cunha, A. F. Precursors' Order Effect on the Properties of Sulfurized Cu2ZnSnS4 Thin Films. *Semicond. Sci. Technol.* **2009**, *24* (10), 105013.

(52) Tanaka, T.; Kawasaki, D.; Nishio, M.; Guo, Q.; Ogawa, H. Fabrication of Cu2ZnSnS4 Thin Films by Co-Evaporation. *Phys. status solidi C* **2006**, *3* (8), 2844–2847.

(53) Hohenberg, P.; Kohn, W. Inhomogeneous Electron Gas. *Phys. Rev.* **1964**, *136* (3B), B864.

(54) Kohn, W.; Sham, L. J. Self-Consistent Equations Including Exchange and Correlation Effects. *Phys. Rev.* **1965**, *140* (4A), A1133.

(55) Kresse, G.; Hafner, J. Ab Initio Molecular-Dynamics Simulation of the Liquid-Metal--Amorphous-Semiconductor Transition in Germanium. *Phys. Rev. B* **1994**, *49* (20), 14251.

(56) Kresse, G.; Furthmüller, J. Efficiency of Ab-Initio Total Energy Calculations for Metals and Semiconductors Using a Plane-Wave Basis Set. *Comput. Mater. Sci.* **1996**, *6* (1), 15–50.

(57) Perdew, J. P.; Burke, K.; Ernzerhof, M. Generalized Gradient Approximation Made Simple. *Phys. Rev. Lett.* **1996**, *77* (18), 3865.

(58) Blöchl, P. E. Projector Augmented-Wave Method. *Phys. Rev. B* **1994**, *50* (24), 17953.

(59) Kresse, G.; Joubert, D. From Ultrasoft Pseudopotentials to the Projector Augmented-Wave Method. *Phys. Rev. B* **1999**, *59* (3), 1758.

(60) Monkhorst, H. J.; Pack, J. D. Special Points for Brillouin-Zone Integrations. *Phys. Rev. B* **1976**, *13* (12), 5188.

(61) Henkelman, G.; Uberuaga, B. P.; Jónsson, H. A Climbing Image Nudged Elastic Band Method for Finding Saddle Points and Minimum Energy Paths. *J. Chem. Phys.* **2000**, *113* (22), 9901–9904.

(62) Freysoldt, C.; Grabowski, B.; Hickel, T.; Neugebauer, J.; Kresse, G.; Janotti, A.; de Walle, C. G. First-Principles Calculations for Point Defects in Solids. *Rev. Mod. Phys.* **2014**, *86* (1), 253.

(63) Zhang, J.; Zhang, Y.; Tse, K.; Deng, B.; Xu, H.; Zhu, J. New Approaches for Calculating Absolute Surface Energies of Wurtzite (0001)/(000 1{\={}}): A Study of ZnO and GaN. *J. Appl. Phys.* **2016**, *119* (20), 205302.

(64) Zhang, Y.; Evans, J. R. G.; Yang, S. Corrected Values for Boiling Points and Enthalpies of Vaporization of Elements in Handbooks. *J. Chem. Eng. Data* **2011**, *56* (2), 328–337.

(65) Wiberg, E.; Holleman, A. F.; Wiberg, N. *Inorganic Chemistry*; Academic press, 2001.